\documentclass[floatfix,aps,pre,twocolumn,showpacs]{revtex4}

\usepackage{graphicx}
\usepackage{amsmath}
\usepackage{amssymb}
\usepackage{mathrsfs}

\begin{document}

\title{Approaching the ground states of the random
maximum two-satisfiability problem by a greedy single-spin flipping process}

\author{Hui Ma and Haijun Zhou}

\affiliation{Key Laboratory of Frontiers in Theoretical Physics and
	    Kavli Institute for Theoretical Physics China, 
	    Institute of Theoretical Physics, Chinese Academy of Sciences,
		Beijing 100190,	China}

\begin{abstract}
In this brief report we explore the energy landscapes of
two spin glass models using
a greedy single-spin flipping process, {\tt Gmax}.
The ground-state energy density of the random
maximum two-satisfiability problem is efficiently approached
by {\tt Gmax}. The achieved energy density
$e(t)$ decreases with the evolution time $t$ as
$e(t)-e(\infty)=h (\log_{10} t)^{-z}$ 
with a small prefactor $h$ and a scaling coefficient $z > 1$, indicating
an energy landscape with deep and rugged funnel-shape regions. For the
$\pm J$ Viana-Bray
spin glass model, however, the greedy single-spin dynamics quickly gets
trapped to a local minimal region of the energy landscape.
\end{abstract}


\maketitle

%%%%%%%
%%%%%%% Introduction
%%%%%%%

There are extensively many competing interactions in a spin glass or a
discrete combinatorial optimization problem. These competing interactions
cause strong
frustrations among the spin values of different variables, leading to a
complex energy landscape. The energy landscape
structure has a fundamental influence on the
equilibrium and dynamical properties of a thermodynamical system,
and it has been
extensively studied in the fields of protein folding
\cite{Onuchic-Wolynes-2004} and supercooled liquids
\cite{Debenedetti-Stillinger-2001}.
But to work out the detailed energy landscapes for some representative models
has been very difficult, and numerical approaches are limited to systems
with $10^2$-$10^3$ particles
(see, e.g., \cite{Kim-Lee-Lee-2007,Carmi-etal-2009}). 
Some properties of the energy landscapes of mean-field spin glasses were
investigated using the replica and the cavity method of statistical mechanics
\cite{Mezard-Montanari-2009}.
These properties include (1) the energy level at which the equilibrium
configuration space
starts to split into exponentially many ergodic subspaces (referred
to as Gibbs states),
(2) the energy level at which a subexponential  number of Gibbs states start
to dominate the equilibrium configuration space, and 
(3) the ground-state energy density. But we know
little about the distribution of minimal energies and the distribution of
energy barrier heights \cite{DallAsta-Pin-Ramezanpour-2009}.
Great efforts were devoted to the special case of 
the ground-states being unfrustrated, for which we enjoy a rather complete
knowledge on the evolution of the ground-state configuration space
\cite{Mezard-etal-2002,Mezard-Zecchina-2002,Achlioptas-RicciTersenghi-2006,Krzakala-etal-PNAS-2007,Zhou-Wang-2010}. Efficient stochastic search algorithms
were designed to construct unfrustrated spin configurations (see, e.g.,
\cite{Mezard-etal-2002,Alava-etal-2008,Gomes-etal-2008}).
Very recently, the cavity method of statistical physics was extended to
study the evolution
of a single Gibbs state with temperature \cite{Zdeborova-Krzakala-2010}.

In this brief report we explore the energy landscapes of two mean-field
spin glass models using
stochastic local dynamics. For the random maximum $2$-satisfiability
(Max-$2$-SAT) problem,
we find that although the
system is in the spin glass phase at low temperatures, its ground-state
energy density (GSED) can be efficiently approached by a greedy
(zero-temperature) single-spin
flipping process {\tt Gmax}. The achieved energy density $e(t)$ by {\tt Gmax}
decreases with the evolution time $t$ as $e(t)= e_0 + h (\log_{10} t)^{-z}$,
with a small prefactor $h$ and a scaling coefficient $z > 1$. The
asymptotic value
$e_0$ is extremely close to the lower bound of mean GSED as calculated by
the first-step replica-symmetry-broken (1RSB) mean-field theory
\cite{Mezard-Parisi-2003}.
These results are quite surprising, as we
anticipated that the ground-states of a spin glass system can only be
approached by
sophisticated processes (such as simulated annealing
\cite{Kirkpatrick-etal-1983},
exchange Monte Carlo \cite{Hukushima-Nemoto-1996}, extremal optimization
\cite{Boettcher-Percus-2001}) but not by a simple greedy local dynamics.
The logarithmic decaying dynamics indicates that {\tt Gmax} is
exploring rugged funnel-shape regions of the energy landscape, whose 
bottom energy densities being very close to the
ground-state energy density.
Quite different dynamical behavior is observed for the $\pm J$ Viana-Bray
model, which is equivalent
to a modified random Max-$2$-SAT problem with short loops.
 This work suggests that
the energy landscapes of different spin glass systems may have qualitatively
different statistical properties.

%%%%%%%
%%%%%%% Mean-field theory on 2-SAT
%%%%%%%

{\em Max-$2$-SAT and Mean-field results}.---A $2$-SAT formula can be represented as
a bipartite graph in which $N$ binary variables $i$ (with spin
$\sigma_i=\pm 1$) are constrained by $M$
clauses $a$. Each clause $a$ is connected to two variables
(say $i$ and $j$) by edges of fixed binary coupling constants
$J_a^i$ and $J_a^j$; its energy is zero (clause satisfied)
if $\sigma_i=J_a^i$ or $\sigma_j=J_a^j$, otherwise its energy is
unity (clause violated). The total energy of a spin configuration
$\{\sigma_i\}$ is
\begin{equation}
	\label{eq:TwoSATenergy}
	E_{2sat}(\sigma_1, \sigma_2, \ldots, \sigma_N) 
	=  \sum\limits_{a=1}^{M}
	 \frac{(1-J_a^i \sigma_i) (1-J_a^j \sigma_j)}{4} \ .
\end{equation}
The energy density is defined as the configuration energy divided by variable
 number $N$.
Constructing spin configurations of the global minimum energy for this model
(the Max-$2$-SAT problem) is
a NP-hard computational task (for work on approximate algorithms, see
\cite{Lewin-Livnat-Zwick-2002} and references therein).
We focus on the ensemble of random $2$-SAT formulas. 
In a random $2$-SAT formula, the two direct neighbors of each
clause are chosen uniformly at random from the $N$ variables, and the quenched
coupling $J_a^i$ between a clause $a$ and a variable $i$ takes $ \pm 1$
values with equal probability. The ground-state
energy density of the random Max-$2$-SAT problem, as a function of the
clause density $\alpha\equiv M/N$, can be estimated theoretically,
giving us an opportunity to quantitatively
measure the performance of a heuristic algorithm.

%
% figure 00
%
\begin{figure}
	\begin{center}
	\includegraphics[width=0.48\textwidth]{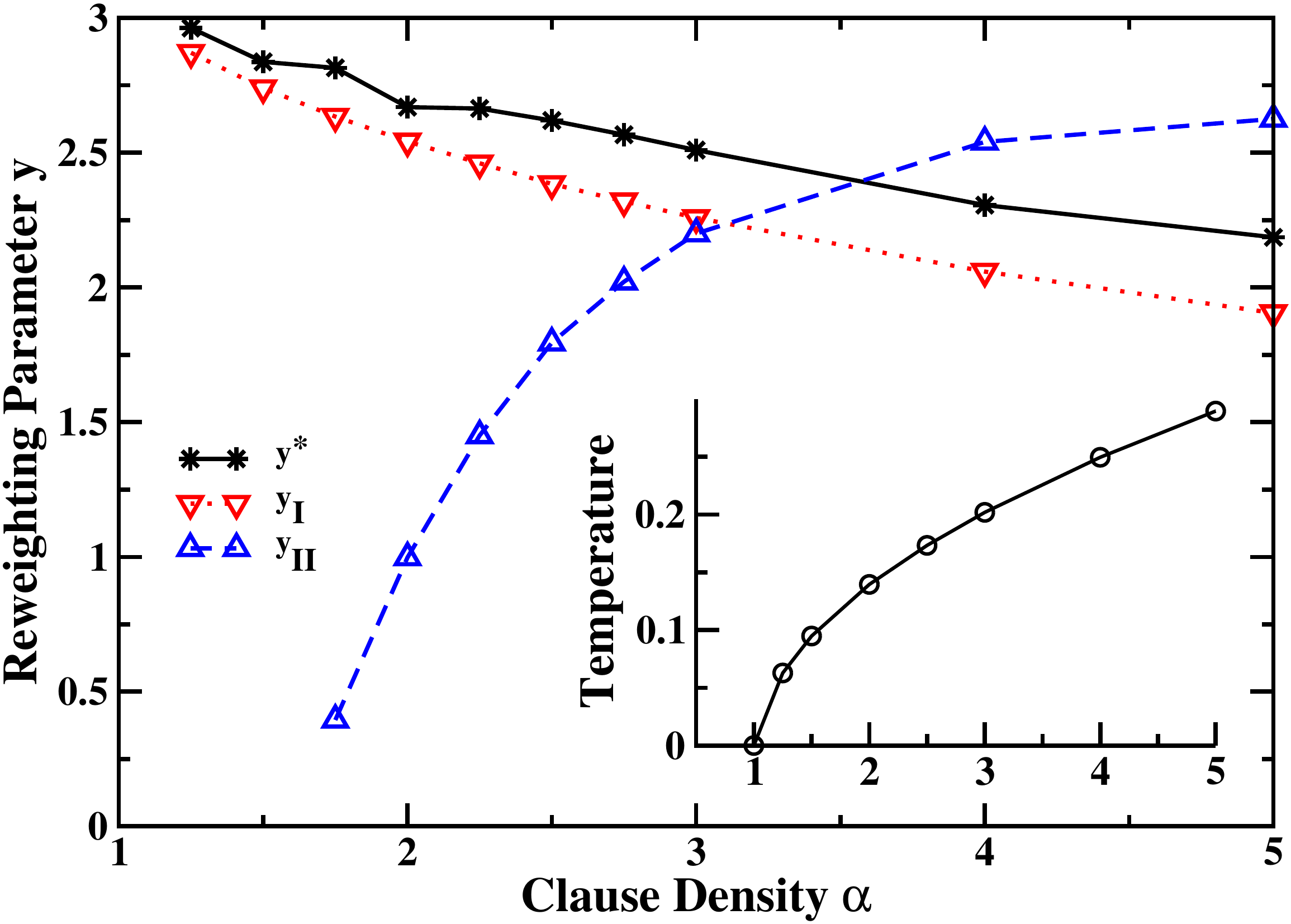}
	\end{center}
	\caption{\label{fig:stability}
	(Color online)
	Stability analysis on the $T=0$ 1RSB mean-field theory for the 
	random Max-$2$-SAT problem.
	$y_I$ and $y_{II}$ are, respectively, the type-I and type-II stability
	threshold for the reweighting parameter $y$.
	$y^*$ is the value of $y$ at which the GSED is
	calculated. The 1RSB theory is stable at $y=y^*$ only if
	$y_{II} < y^* < y_{I}$. The inset shows the
	spin glass transition temperature $T_s$.
        }
\end{figure}

A phase transition occurs for the random Max-$2$-SAT problem at $\alpha=1$,
where the mean GSED starts to be positive. The system is then in
the spin glass phase at low temperatures for $\alpha>1$.
We have determined the spin glass transition temperature $T_s(\alpha)$
using the 1RSB mean-field theory
\cite{Mezard-Parisi-2001,Montanari-etal-2008} (see inset
of Fig.~\ref{fig:stability}). At temperature $T=T_s(\alpha)$, the
equilibrium configuration space divides into exponentially many
Gibbs states.
The mean GSED is calculated within the 1RSB theory by
letting $T=0$ and weighting each
Gibbs state by a factor $e^{-y E_\gamma^m}$, where $E_\gamma^m$ is
the minimum energy of Gibbs state $\gamma$ and the reweighting
parameter $y$ is set to a particular value $y=y^*$
\cite{Mezard-Parisi-2003}. At each value of clause density $\alpha$,
the 1RSB mean-field theory has two stability thresholds $y_I$ and $y_{II}$
\cite{Montanari-etal-2004,Mertens-etal-2006}; for the mean-field theory to be
stable at $y=y^*$, it is required that $y_{II} < y^* < y_{I}$. As we show
in Fig.~\ref{fig:stability}, this condition is violated at $\alpha\geq 1$,
suggesting that the GSED obtained by the
1RSB theory is a lower-bound of the true GSED \cite{Franz-Leone-2003}.

%%%%%%%
%%%%%%% Gmax algorithm and simulatin results
%%%%%%%

{\em Greedy single-spin flipping for Max-$2$-SAT}.---A single-spin flipping process,
 {\tt Gmax}, is used to construct low-energy spin configurations
for single $2$-SAT formulas. The process starts from a random initial
spin configuration $\{\sigma_i(0)\}$ at evolution time $t=0$.
The evolution time then increases with step $\Delta t=1/N$.
A spin configuration $\{ \sigma_i(t)\}$ at time $t$ is associated with a
candidate variable set $C(t)$. A variable belongs to  $C(t)$
if and only if reversing
its spin does not cause an increase of
the configuration energy.
A variable $i$ is chosen from the set $C(t)$ uniformly at random, and at
evolution time $t^\prime = t+\Delta t$, the spin configuration is
updated to $\{\sigma_i(t^\prime)\}$ which differs from $\{\sigma_i(t)\}$ only at
variable $i$. After this spin flipping,
the candidate set is also updated to $C(t^\prime)$.

%
% figure evolution trajectory
%
\begin{figure}
	\begin{center}
	\includegraphics[width=0.48\textwidth]{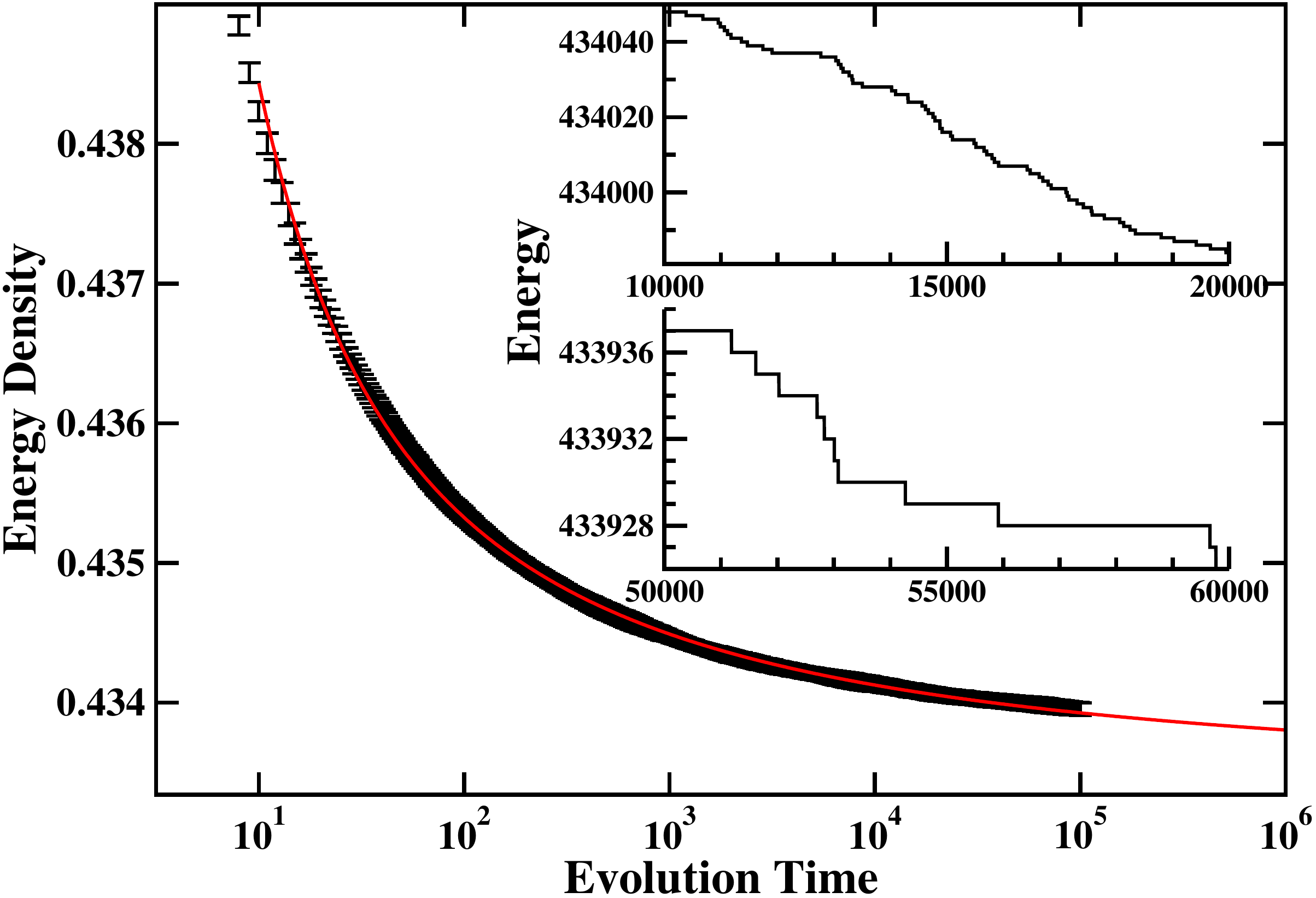}
	\end{center}
	\caption{\label{fig:EnergyEvolution}
	(Color online)
	The energy density reached by {\tt Gmax} as a function of
	evolution time $t$. The energy trajectory is an averaging
	over $16$ repeated running of {\tt Gmax} on a single
	random $2$-SAT formulas of size $N=10^6$ and
	$\alpha=5.0$. The solid line is the fitting curve of the
	form $e(t)=e_0 + h/(\lg t)^z$, with parameters
	$(e_0, h, z)$ being $(0.433372(8), 0.005063(8), 1.375(7))$.
	The asymptotic value $e_0$ is very close to the 
	1RSB lower bound of $0.43273$.
	Inset shows two segments of a single evolution trajectory.
	}
\end{figure}

{\tt Gmax} is a greedy process, the configuration energy never increases
 with $t$.
The density $e(t)$ of energy at time $t$ is
$e(t)\equiv E(\sigma_1(t), \sigma_2(t),\ldots, \sigma_N(t) )/N$.
As an example, Fig.~\ref{fig:EnergyEvolution} shows the evolution of $e(t)$
for a random $2$-SAT formula with
$N=10^6$ variables and clause density $\alpha=5.0$. We notice that initially $e(t)$ decreases
very fast with $t$. As $t$ increases, the decreasing rate
becomes slower and slower and the time interval between two consecutive
energy decreases becomes longer and longer
(inset of Fig.~\ref{fig:EnergyEvolution}).
However, the energy of a single trajectory keeps decreasing even for
$t > 10^5$, suggesting that {\tt Gmax} is not being trapped by
a local energy minimum.   The decreasing of $e(t)$ can be
well fitted by the following form
\begin{equation}
e(t) = e_0 + \frac{h}{(\log_{10} t)^z} \ ,
\label{eq:EnergyEvolution}
\end{equation}
with parameters $e_0$, $h$, and $z$. The value $e_0$ is the asymptotic value
of $e(t)$ at $t\rightarrow \infty$, its value is slightly different for
different random
$2$-SAT instances with the same $N$ and $\alpha$.
The fitting parameter $h\sim 10^{-3}$ is much
smaller than $e_0$, and the scaling exponent $z$ is larger than unity.
 The observations, 
(i) $e(t)$ can be expressed as a function of $\log(t)$, and (ii)
the asymptotic value $e_0$ is very close to the 1RSB lower bound of the mean
GSED, suggest that the random Max-$2$-SAT problem has a
funnel-shaped energy landscape, to some extent similar to those of protein
sequences \cite{Onuchic-Wolynes-2004}. {\tt Gmax} visits different
funnel-shaped regions when repeated on the same $2$-SAT formula
starting from different initial spin configurations, and the bottom energies of
these different funnels are very close to each other. However, if
{\tt Gmax} is made to be more greedy, e.g., by preferentially flipping
those spins that lead to an energy decrease, the modified dynamics is
usually trapped by local minimal regions of the energy landscape.

%
% figure energy density
%
\begin{figure}
	\begin{center}
	\includegraphics[width=0.48\textwidth]{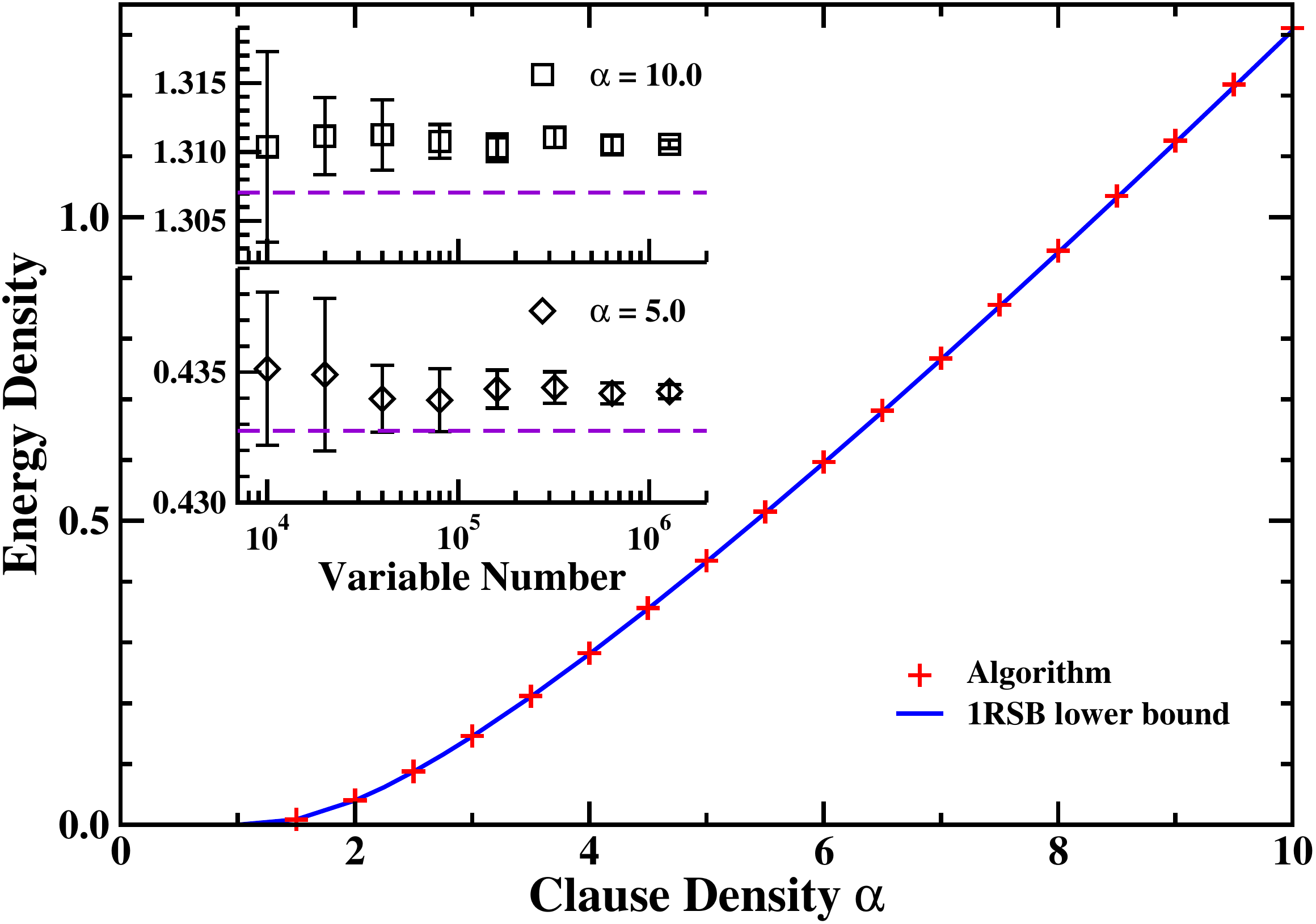}
	\end{center}
	\caption{\label{fig:EnergyAlphaNumber}
	(Color online)
	Comparison between the energy density reached by a single run of
	{\tt Gmax} on a random $2$-SAT formula of $N=10^6$ variables
	and $\alpha N$ clauses (crosses) and the 1RSB lower bound of mean
	GSED (solid line). The inset shows the mean energy
	density reached by a single run of {\tt Gmax}
	 on $16$ random $2$-SAT formula of fixed clause density $\alpha=5.0$
	(diamonds) and $\alpha=10.0$ (squares) and different variable
	numbers $N$, with the dashed lines marking the
	corresponding 1RSB lower bounds.
	 The waiting time of {\tt Gmax}
	is set to $t=10^4$ in all these simulations.
	}
\end{figure}

Figure~\ref{fig:EnergyAlphaNumber} shows the energy density reached by
a single run of {\tt Gmax} at evolution time $t=10^4$ on a random $2$-SAT
formula of $N=10^6$ variables and clause density $\alpha \in [1,10]$.
The simulation results are in excellent agreement with the energy density
lower bound as obtained by the
1RSB mean-field theory, indicating that {\tt Gmax} is able to approach the
GSED of a large random $2$-SAT formula within a reasonable waiting time. As a
stochastic greedy local search process, {\tt Gmax} is also very fast.
We have compared the performance of {\tt Gmax} with that of the message-passing
algorithm {\tt SP-y}  on several 
large random $2$-SAT problem instances with $N=10^6$ and $\alpha=5$ and $\alpha=10$.
{\tt Gmax} is about ten times
faster than {\tt SP-y} and it reaches noticeably lower energy values.
The global algorithm {\tt SP-y} is inspired by
the 1RSB spin glass theory \cite{Mezard-Montanari-2009,Mezard-etal-2002,Mezard-Zecchina-2002};
it contains several adjustable parameters, including the
reweighting parameter $y$. As the 1RSB mean-field theory is not sufficient for the random
$2$-SAT problem, the message-passing routine of {\tt SP-y} does not converge. One has to run
the non-convergent {\tt SP-y} process many times with different $y$ values to get good results.
The best results obtained by repeated running of {\tt SP-y} are typically worse than the result
of a single run of {\tt Gmax}.

The inset of Fig.~\ref{fig:EnergyAlphaNumber} shows the mean energy density value reached by
a single run of {\tt Gmax} at evolution time $t=10^4$ on random $2$-SAT formulas of
fixed clause density $\alpha$ and different sizes $N$. The reached
energy density has only a tiny gap of about $0.002$ above the 1RSB lower bound at $t=10^4$,
and this gap does not increase with $N$. Such a tiny gap can be further
reduced simply by waiting longer.

{\em {\tt Gmax} for Viana-Bray model}.---The $\pm J$ Viana-Bray model on random graphs is
equivalent to the random $2$-exclusive-or-satisfiability ($2$-XORSAT) problem, its
energy function can be written as
\begin{equation}
	\label{eq:2xorsat}
	E_{2xor}(\sigma_1, \sigma_2, \ldots, \sigma_N) =
	 \sum\limits_{a=1}^{M} \frac{1-J_a \sigma_i \sigma_j}{2} \ .
\end{equation}
The coupling constant $J_a$ of a clause $a$
takes $\pm 1$ values with equal probability, and the involved two variables $i$ and
$j$ of a clause $a$ are chosen from the whole set of $N$ variables uniformly at
random. The ground-state energy density of Eq.~(\ref{eq:2xorsat}) is positive
at clause density $\alpha\equiv M/N>0.5$. For $\alpha>0.5$, the random $2$-XORSAT is
in the spin glass phase at low temperatures.  A lower bound for the GSED
can be obtained using the zero-temperature 1RSB mean-field theory.

%
% figure energy density 2-xorsat
%
\begin{figure}
	\begin{center}
	\includegraphics[width=0.48\textwidth]{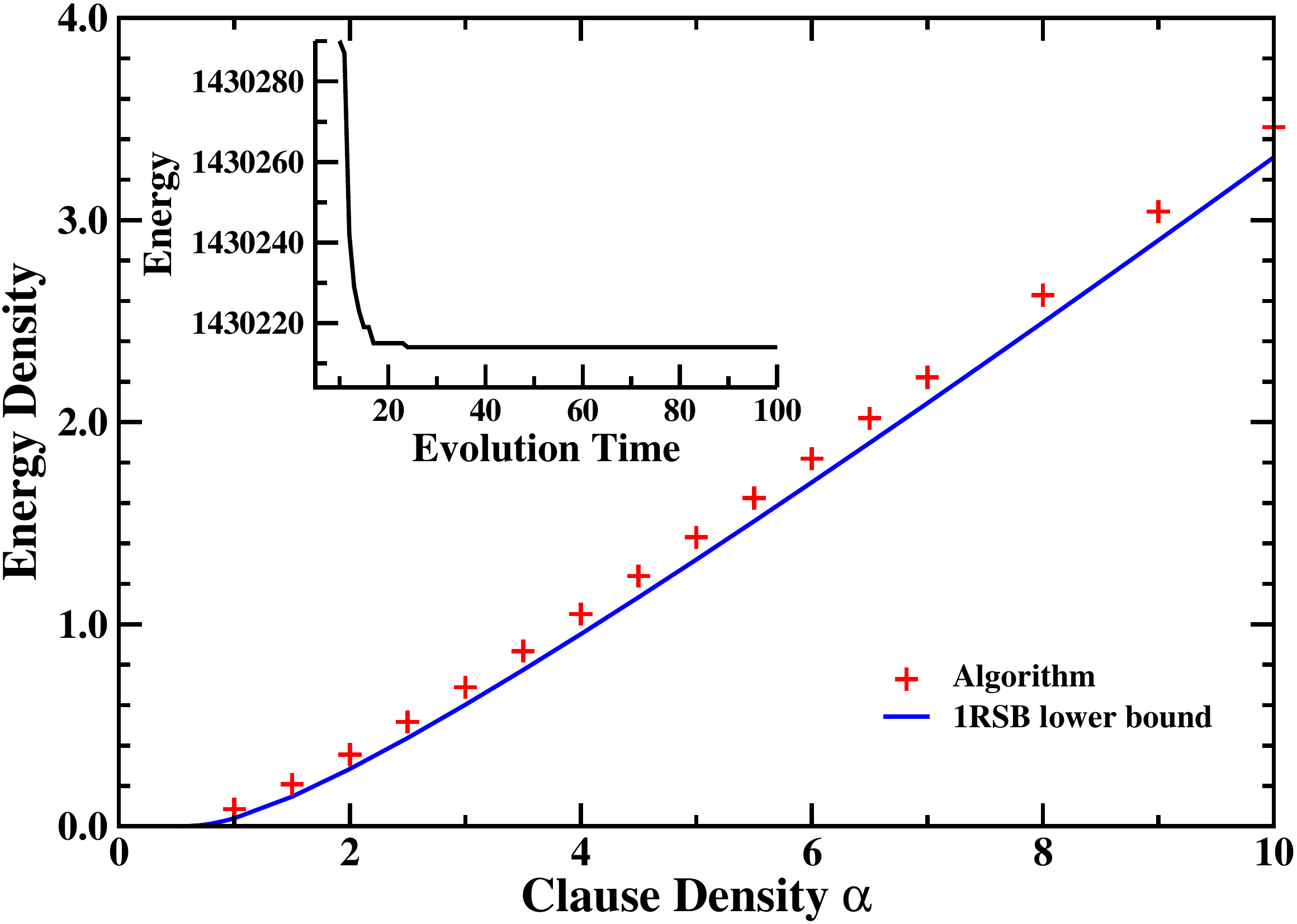}
	\end{center}
	\caption{\label{fig:EnergyAlphaXORSAT}
	(Color online)
	Comparison between the energy density reached by a single run of
	{\tt Gmax} on a random $2$-XORSAT formula of $N=10^6$ variables
	and $\alpha N$ clauses (crosses) and the 1RSB lower bound of mean
	GSED (solid line). The inset shows the
	evolution trajectory of a single run of {\tt Gmax} on a problem
	instance with $N=10^6$ and $\alpha=5.0$.
	}
\end{figure}

Figure ~\ref{fig:EnergyAlphaXORSAT} compares the results of a single run of
{\tt Gmax} on a random $2$-XORSAT formula with $N=10^6$ variables and
$M= \alpha N$ clauses and the energy density 1RSB lower bound.
We notice that (see inset of Fig.~\ref{fig:EnergyAlphaXORSAT}),
{\tt Gmax} quickly comes to a local minimum region of the energy
landscape and it is then not able to escape. This is
dramatically different from the dynamical behavior observed for the random
Max-$2$-SAT problem. There is a noticeable
gap between the energy density reached by {\tt Gmax} and the
1RSB lower-bound.

A $2$-XORSAT constraint $a$ can be expressed in terms of two
$2$-SAT constraints: $(1-J_a \sigma_i \sigma_j)/2 \equiv  
(1 - \sigma_i) (1+J_a \sigma_j)/4+ (1+\sigma_i) (1-J_a \sigma_j)/4$.
These two SAT constraints form a short loop between 
the two involved variables $i$ and $j$. 
Therefore, a random $2$-XORSAT formula with clause density
$\alpha$ can be converted to a modified random $2$-SAT formula with
clause density $2 \alpha$. 
The GSED of the random
$2$-XORSAT problem at clause density $\alpha$ is only slightly
higher than that of the random $2$-SAT problem at clause density
$2 \alpha$ (for example, according to the 1RSB mean-field theory,
the GSED of random $2$-XORSAT is
$1.320$ at $\alpha=5.0$ and that of random $2$-SAT
is $1.307$ at $\alpha=10$).
The existence of an extensive number of short loops in the modified
random $2$-SAT
problem therefore does not have much influence on the GSED.
However, it changes the system's energy landscape significantly.

%%%%%%%
%%%%%%% Discussion and conclusion
%%%%%%%

{\em Discussion}.---The random Max-$2$-SAT problem is in the spin glass phase
at low temperatures. However we found that, for single large Max-$2$-SAT
problem instances, a simple greedy single-spin flipping process
{\tt Gmax} is able to reach spin configurations with
energy densities extremely close to the lower bound of mean ground-state energy
density as predicted by the 1RSB mean-field theory.
Such an observation is contrary to the conventional belief that
greedy dynamics will be trapped into local stable regions of the energy
landscape, whose energy densities are markedly higher than the ground-state
value \cite{RicciTersenghi-2010}.
From the logarithmic dependence of the energy density $e(t)$ with
the evolution time $t$ [Eq.~(\ref{eq:EnergyEvolution})], we infer that, as
the reached energy density is close to the asymptotic value $e_0$, the
configuration energy is further decreased by accumulating spin local
modifications into configuration rearrangements of larger and
larger scale \cite{Zhou-Wang-2010}. This dynamical behavior is consistent
with an energy landscape with many
rugged and deep funnel-shaped regions. On the other hand,
the energy landscape
of the Max-$2$-XORSAT problem (the Viana-Bray model) appears to have
a very different structure.

Our findings call for understanding from the theoretical side.
{\tt Gmax} is simply a single-spin Glauber dynamics quenched at
zero temperature. It is highly
desirable to have a theoretical understanding on
the empirical observation Eq.~(\ref{eq:EnergyEvolution}). At the
spin glass transition temperature $T_s$, many Gibbs states
start to form in the equilibrium configuration space. Suppose they
are uniformly sampled, what is their energy depth distribution?
And how will this distribution change if the Gibbs states at $T_s$ are sampled
according to the Boltzmann distribution?
The theoretical framework of \cite{Zdeborova-Krzakala-2010} may be useful for
answering these questions.

For random Max-$K$-SAT problems with $K\geq 3$, our simulation results (not
shown) suggest that {\tt Gmax} is not able
to find spin configurations with energy densities extremely close to
the 1RSB-predicted ground-state value. However, the gap between the reached
energy density and the ground-state energy density is still very small 
(for example, {\tt Gmax} reaches an
energy density of $0.3315$ at $10^4$ evolution steps for a random $3$-SAT
formula of $10^6$ variables and $\alpha=10$, close to
the GSED value of $0.3114$ by 1RSB theory;
for a single random $4$-SAT formula of $10^6$ variables
and $\alpha=20$, these two energy densities are $0.2960$ versus $0.2493$).

%%%%%%% 
%%%%%%%   Acknowledgement
%%%%%%%

This work was partially supported by NSFC Grants 10774150, 10834014, and
the 973-Program Grant 2007CB935903.
Simulations were performed on the HPC computer cluster of ITP-CAS.

%%%%%%%   References

%\bibliography{/cygdrive/d/ResearchPapers/references}

\end{document}